\documentclass{iaus}
\usepackage{graphicx}

\title[Disk in X Her] 
{Disklike Structure in the Semiregular Pulsating Star X Her}

\author[Nakashima, J.]   
{Jun-ichi Nakashima$^{1,2}$%
  }

\affiliation{$^1$Institute of Astronomy and Astrophysics, 
Academia Sinica, P.O. Box 23-141, Taipei 106, Taiwan \\[\affilskip]
$^2$Department of Astronomy, University of Illinois at Urbana-Champaign,
1002, W. Green St., Urbana, IL 61801, USA \break email: junichi@asiaa.sinica.edu.tw}

\pubyear{2006}
\volume{234}  
\pagerange{xxx--xxx}
\date{?? and in revised form ??}
\jname{Proceedings Title IAU Symposium}
\editors{A.C. Editor, B.D. Editor \& C.E. Editor, eds.}
\begin{document}

\maketitle

\begin{abstract}
The author presents the results of Berkeley-Illinois-Maryland (BIMA) array interferometric observations in the CO $J=1$--0 line toward X Her and EP Aqr, the semiregular pulsating stars with a composite CO line profile, and also reports finding of a disklike structure in X Her. In the CO spectrum both of X Her and EP Aqr, a composite profile including narrow and broad components is seen as reported by the previous single-dish observations. The spatial structure of the broad component region of X Her shows a bipolar shape, and that of the narrow component shows an elliptical/spherical shape. The blue- and red-shifted parts of the X Her narrow component show a systematic difference in the velocity integrated intensity map. The spatio-kinetic properties of the X Her narrow component are reminiscent of a Keplerian rotating disk with a central mass of 0.9 M$_{\odot}$. The spatial distributions of both the narrow and the broad components of EP Aqr appear to be roughly round with the same peak positions; no significant velocity gradient is seen. The spatio-kinetic properties of EP Aqr are reminiscent of a multiple-shell structure model rather than of a bipolar flow and disk model. 
\end{abstract}

\firstsection
\section*{}
Asymptotic giant branch (AGB) stars occasionally exhibit a molecular line profile with very small line widths less than 5 km s$^{-1}$, much smaller than a typical AGB outflow velocities. Such a narrow line profile is usually superimposed on a broad pedestal component (this profile is often called a ``double component profile'', or a ``composite profile''), but is sometimes independently found without the broad component. A non-negligible number of AGB stars exhibit a narrow line; in fact, the narrow line has been detected from 5--10\% of a sample of AGB stars. Narrow line profiles have been detected toward a wide variety of AGB stars. AGB stars exhibiting a narrow line profile are notable in some respects. First, chemically unusual AGB stars occasionally exhibit a narrow line profile. For instance, Kahane \& Jura (1996) reported double component profiles in the CO radio lines of BM Gem and EU And; these stars are ``silicate carbon stars,'' which simultaneously have a carbon-rich (C-rich) central star and oxygen-rich (O-rich) circumstellar material. Nakashima \& Deguchi (2004) have found a double component profile in radio molecular spectra of the SiO maser source with a rich set of molecular species, IRAS 19312+1950 (Nakashima \& Deguchi 2005). Second, according to recent observations, some of AGB stars exhibiting a narrow line profile might harbor a Keplerian rotation disk. Bergman et al. (2000) have reported a tentative detection of such a disk in the O-rich AGB star with a double component profile, RV Boo. Though a physical relationship between narrow lines, chemical peculiarity and a Keplerian rotation disk is still unclear, Kahane \& Jura (1996) have suggested that narrow lines found in silicate carbon stars might be explained by gravitationally stable material in the form of a distorted or puffed-up slowly rotating disk, in which O-rich material is trapped. In contrast, to explain a double component profile, Knapp et al. (1996) have advanced a hypothesis that takes multiple shell structure into account, in which each shell has different expanding velocities produced by episodic mass loss with highly varying gas expansion velocities. 

EP Aqr and X Her are a semiregular pulsating O-rich AGB star with a double component CO line profile. Interferometric observations of these two stars in the CO $J=1$--0 line were made with the BIMA array from 2004 January to May (see Nakashima 2005, 2006 in details of the observations). Fig. 1. shows a spatially averaged spectra and velocity integrated intensity maps. At a glance, the properties of X Her and EP Aqr seen in the intensity map are very different. X Her seems to have a bipolar flow (high velocity component) and disklike component (low velocity component), and EP Aqr seems to have two spherically expanding shells (high and low velocity components). The disklike component seen in the X Her map appears to exhibit rotating motion; the best fit results of the Keplerian function to the data gives a central mass of about 0.9 M$_{\odot}$. In contrast, the most natural interpretation for EP Aqr would be a double shell with different expanding velocity if we do not see a bipolar flow and disk in the pole on view.

\begin{figure}
\centerline{
\scalebox{0.57}{%
\includegraphics{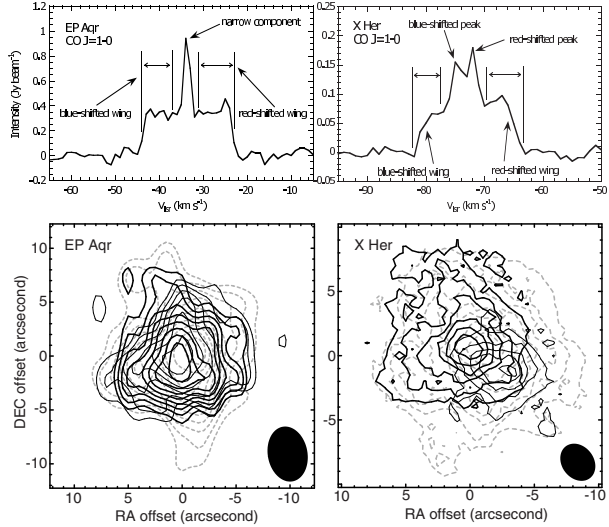}%
}
}
  \caption{{\it Upper panels}: Spatially averaged spectra of X Her and EP Aqr in the CO $J=1$--0 line. The averaging area is a circle with a diameter of 15$''$. {\it Lower panels}: Velocity integrated intensity maps of EP Aqr and X Her. The thick and thin contours map the blue- and redshifted wings of the broad component, respectively; the gray dashed contours map the narrow component. The synthesized beam size is indicated in the bottom right corner. The contours start at the 5$\sigma$ level.}\label{fig:sp-and-map}
\end{figure}

\end{document}